\documentclass[twocolumn,twocolappendix]{aastex631}

\usepackage{graphicx,enumitem}
\usepackage{xcolor,graphicx,xspace,color,longtable,}
\usepackage{amssymb,amsmath,shadow,bezier,curves,rotating}
\usepackage{graphicx,chngcntr}
\graphicspath{ {./images/}}
\usepackage{color,soul}
\usepackage{booktabs}
\usepackage[flushleft]{threeparttable}
\usepackage[autostyle=true]{csquotes}
\usepackage[english]{babel}
\usepackage{hyperref}
\usepackage{tabularx} 
\setlist[enumerate]{leftmargin=0pt,labelindent=0pt}

\newcommand {\h}    {$h^{-1}\,\mathrm{Mpc}$}

\defcitealias{Biviano16}{Biv16}
\defcitealias{Biviano21}{Biv21}

\DeclareSymbolFont{matha}{OML}{txmi}{m}{it}
\DeclareMathSymbol{\varv}{\mathord}{matha}{118}
\begin{document}

\author[0000-0003-3595-7147]{Mohamed H. Abdullah}
\affiliation{Department of Physics, University of California Merced, 5200 North Lake Road, Merced, CA 95343, USA}
\affiliation{Department of Astronomy, National Research Institute of Astronomy and Geophysics, Cairo, 11421, Egypt}

\author[0009-0004-4674-6446]{S. W. El-Sogheir}
\affiliation{Department of Astronomy, Space Science, and Meteorology, Faculty of Science, Cairo University, Giza 11326, Egypt}

\author[0000-0002-6572-7089]{Gillian Wilson}
\affiliation{Department of Physics, University of California Merced, 5200 North Lake Road, Merced, CA 95343, USA}

\author[0009-0008-5349-5410]{Magdy Y. Amin}
\affiliation{Department of Astronomy, Space Science, and Meteorology, Faculty of Science, Cairo University, Giza 11326, Egypt}

\author[0000-0001-9070-4914]{A. Ahmed}
\affiliation{Department of Astronomy, Space Science, and Meteorology, Faculty of Science, Cairo University, Giza 11326, Egypt}
%%%%%%%%%%%%%%%%%%% TITLE PAGE %%%%%%%%%%%%%%%%%%%
\title[From Cluster Core to Splashback: Linking Dynamical Structure to Multidimensional Galaxy Evolution in the Coma Cluster]{From Cluster Core to Splashback: Linking Dynamical Structure to Multidimensional Galaxy Evolution in the Coma Cluster}

\begin{abstract}
We investigate galaxy evolution across the full dynamical structure of the Coma cluster using the $\mathtt{GalWCat19}$ spectroscopic cluster catalog combined with SDSS-based value-added galaxy properties. We measure a splashback radius of $R_{\mathrm{sp}} = 2.94 \pm 0.16~h^{-1}\,\mathrm{Mpc}$. Using specific star formation rate ($\mathrm{sSFR}$), color offset from the red sequence ($\Delta(g-r)_{\mathrm{RS}}$), and bulge-to-total ratio ($B/T$) as independent diagnostics, we find a coherent environmental transition from quiescent, red, bulge-dominated galaxies in the cluster core to increasingly star-forming, blue, disk-dominated populations at larger radii. We further introduce a three-dimensional framework in the joint $(\log \mathrm{sSFR}, \Delta(g-r)_{\mathrm{RS}}, B/T)$ space and develop a peak-based classification scheme that extends beyond traditional one-dimensional galaxy classifications. This framework identifies two dominant populations: red, quiescent, bulge-dominated galaxies, which account for $51\%$ of the joint-analysis sample, and blue, star-forming, disk-dominated galaxies, which account for $29\%$. The remaining $\sim20\%$ of galaxies occupy transitional or mixed states that connect these two principal populations. The relative fractions of these populations change strongly near the splashback radius, where the red, quiescent, bulge-dominated population declines rapidly and the blue, star-forming, disk-dominated population becomes increasingly dominant. These results show that the splashback boundary is not only a dynamical boundary, but also a critical evolutionary transition zone. Overall, our findings suggest that galaxy evolution in Coma is not a purely binary transformation, but instead proceeds through continuous multidimensional pathways in which star formation quenching, color evolution, and morphological transformation occur on different timescales while remaining closely linked to the cluster dynamical structure.
\end{abstract}

%%%%%%%%%%%%%%%%% BODY OF PAPER %%%%%%%%%%%%%%%%%%
\section{Introduction} \label{sec:intro}

The Coma cluster (Abell~1656) is one of the nearest and richest galaxy clusters in the local universe, making it a fundamental benchmark for studies of cluster astrophysics, large-scale structure, and environmental galaxy evolution \citep[e.g.,][]{Kent82,Colless96,Carter08}. Its proximity at $z = 0.0234$ and high mass of $\sim10^{15}\,M_\odot$ \citep[e.g.,][]{Abdullah20a} provide an ideal laboratory for investigating how dense environments influence galaxy evolution \citep[e.g.,][]{Yagi10}. Historically, Coma has played a central role in the development of cluster astrophysics, from early dynamical mass estimates based on velocity dispersion \citep{Zwicky09} to modern studies that test and calibrate mass measurement techniques \citep{Falco14}. Its location within the Coma Supercluster, connected to nearby systems such as Abell~1367 through filamentary structures, further enables studies of environmental effects across a wide range of densities and scales \citep[e.g.,][]{Gavazzi10,Tiwari20}.

Previous studies have investigated a broad range of Coma’s physical and galaxy population properties, including its internal dynamics, substructure, luminosity and stellar mass functions, morphology, star formation activity, and large-scale environment \citep[e.g.,][]{Colless96,Smith12,Sohn17,Kadowaki17}, providing detailed constraints on the structure of the cluster and its galaxy populations. As a well-studied nearby cluster, Coma has played a central role in establishing key observational links between dense environments and galaxy evolution, including morphology--density relations, environmental quenching, red-sequence growth, and the segregation between virialized cores and infall regions \citep[e.g.,][]{Dressler80,Balogh04,Peng10,Boselli14,Poggianti17,Rhee17}. Despite these advances, most studies focus on individual components and do not link them within a unified, multidimensional framework tied to the full dynamical structure of Coma.

The influence of environment on galaxy evolution is one of the central problems in extragalactic astronomy. In dense cluster environments, galaxies are subject to external processes that can significantly alter their star formation activity, stellar populations, gas content, and morphology beyond what is expected from internal secular evolution alone \citep[e.g.,][]{Larson80,Moore96,Boselli06,Ebeling14}. Processes such as ram-pressure stripping, starvation, tidal interactions, harassment, and preprocessing suppress star formation, remove or heat gas, and drive structural transformation \citep[e.g.,][]{Boselli22,Marasco23,Tagliaferro21}. These environmental effects are reflected observationally in the excess of red, quiescent, bulge-dominated galaxies in dense cluster cores and the increasing prevalence of blue, star-forming, disk-dominated systems toward lower-density environments \citep[e.g.,][]{Peng10,Wetzel13}. While these trends demonstrate the strong role of environment in regulating galaxy evolution, many studies have examined star formation, color, or morphology independently using one-dimensional classifications \citep[e.g.,][]{Strateva01,Baldry04,Brinchmann04}. Such approaches can obscure the intrinsically multidimensional nature of galaxy transformation, in which quenching, color evolution, and structural changes may occur on different physical and temporal scales \citep[e.g.,][]{Muzzin12,Wetzel13}.

A major advance in linking cluster dynamics and galaxy evolution is the identification of the splashback radius $R_{\mathrm{sp}}$ as a physically motivated halo boundary \citep[e.g.,][]{Diemer14,Adhikari14,Diemer17}. Unlike $R_{200}$, defined as the radius within which the mean enclosed density is 200 times the critical density of the Universe $\rho_c$,  the splashback radius traces the first apocenter of recently accreted material and more directly reflects a galaxy’s accretion history. This boundary marks the transition between the virialized cluster interior and the infall region \citep[e.g.,][]{More15,Aung23}, providing a physically meaningful division between galaxies that have experienced the full cluster environment and those that are more recently accreted. Both theoretical and observational studies show that $R_{\mathrm{sp}}$ is associated with sharp features in halo density profiles and significant changes in galaxy populations, making it a key scale for linking dynamics to environmental transformation \citep[e.g.,][]{Chang18,Shin19}. By connecting galaxy properties directly to the splashback radius, it becomes possible to investigate where and how environmental processing accelerates as galaxies transition from infall to virialized cluster membership.

Although the Coma cluster has been studied extensively for decades, recent advances in cluster membership identification and galaxy property measurements provide an opportunity for a more precise reexamination of this benchmark system. The FoG-GalWeight technique \citep{Abdullah18,Abdullah20a} enables improved identification of cluster members and projected phase-space boundaries relative to traditional methods, while modern value-added catalogs offer more accurate measurements of galaxy stellar populations and structure. By combining this high-purity membership framework with stellar masses, star formation rates (SFRs), and specific star formation rates (sSFRs) from the MPA/JHU catalog\footnote{\url{http://www.mpa-garching.mpg.de/SDSS/DR7/}}, together with refined photometric and structural measurements from \citet{Meert15}, we construct a self-consistent dataset that links cluster dynamics to galaxy properties. These advances allow us to study Coma within a unified dynamical, stellar population, and morphological framework.

In this work, we investigate galaxy evolution across the full dynamical structure of the Coma cluster, from the virialized core to the splashback boundary and infall region. We examine how galaxy populations vary across this structure using star formation activity, color, and morphology both independently and jointly. To move beyond traditional one-dimensional classifications, we introduce a three-dimensional framework in the joint $(\Delta(g-r)_{\mathrm{RS}}, \log \mathrm{sSFR}, B/T)$ space and develop a peak-based classification scheme that identifies the dominant galaxy populations directly from the intrinsic multidimensional distribution. This approach enables us to quantify the principal evolutionary populations, identify transitional systems, and determine how galaxy transformation from the cluster core to the splashback region is linked to the underlying dynamical structure.

The paper is organized as follows. In Section~\ref{sec:data}, we describe the data, sample selection, and the galaxy physical and structural properties used throughout this study. In Section~\ref{sec:results}, we present our analysis of the Coma cluster, including its projected spatial and dynamical structure, surface density profile and splashback radius, environmental trends in galaxy star formation, color, and morphology, and our new three-dimensional evolutionary framework. Finally, in Section~\ref{sec:conc}, we summarize our main results and conclusions. Throughout this paper, we adopt a flat $\Lambda$CDM cosmology consistent with the \citet{Planck15} results, assuming $\Omega_{\mathrm{M}} = 0.3089$, $\Omega_{\Lambda} = 0.6911$, and $h = 0.6774$. We use the term `log' to refer to base-10 logarithms throughout.
%%%%%%%%%%%%%%%%%%%%%%%%%%%%%%%%%%%%%%%%%%%%%%%%%%%%%%
%Figure 1_______________________________
\begin{figure*}\hspace{0cm}
\centering
\includegraphics[width=1\linewidth]{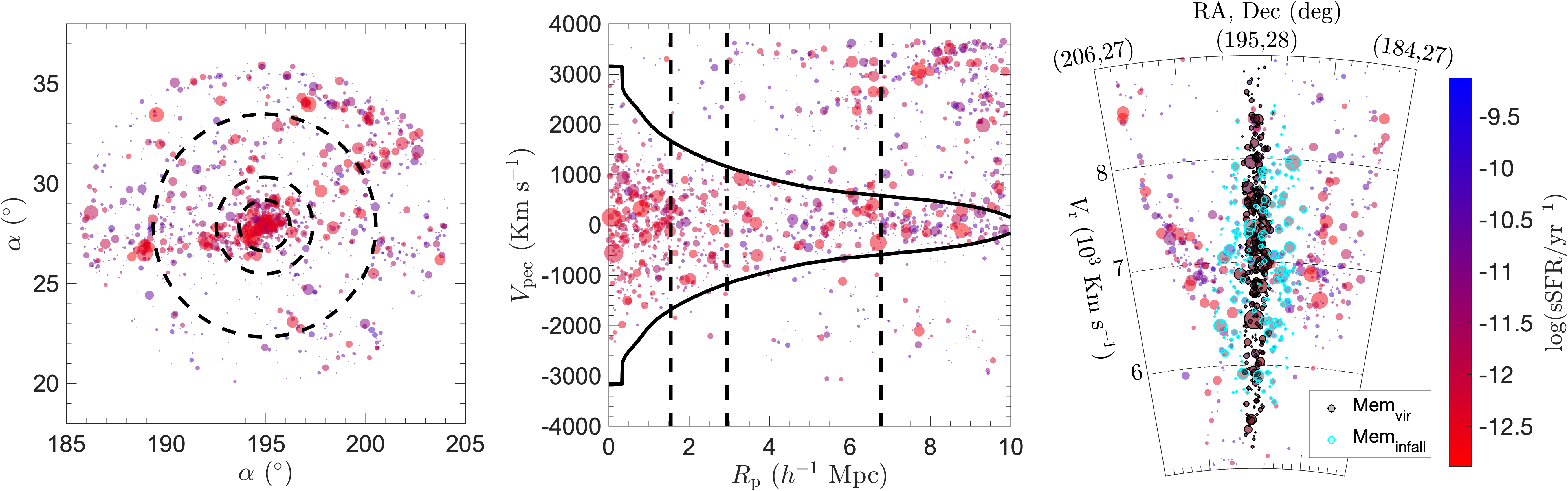} \vspace{-0.5cm}
\caption{
Spatial and projected phase-space distribution of galaxies in the Coma cluster region. 
Left: sky distribution of galaxies centered on Coma. 
Middle: projected phase-space diagram showing the line-of-sight peculiar velocity ($V_{\mathrm{pec}}$) as a function of projected radius ($R_p$). The solid curves show the phase-space boundaries derived from the FoG-GalWeight technique \citep{Abdullah18}. 
Right: redshift-space slice of the galaxy distribution, shown in sky position and radial velocity and projected along the declination direction. 
Galaxies identified as virialized members ($\mathrm{Mem}_{200}$) are highlighted by black circles, while galaxies in the infall region ($\mathrm{Mem}_{\mathrm{infall}}$) are highlighted by cyan circles. 
The dashed circles and vertical dashed lines indicate $R_{200}$, $R_{\mathrm{sp}}$, and $R_{\mathrm{ta}}$, where applicable. 
Galaxies are color-coded by $\log(\mathrm{sSFR}/\mathrm{yr}^{-1})$, and marker sizes scale with stellar mass.
}
\label{fig:sky}
\end{figure*}
%%%%%%%%%%%%%%%%%%%%%%%%%%%%%%%%%%%%%%%%%%%%%%%%%%%%%%
%%%%%%%%%%%%%%%%%%%%%%%%%%%%%%%%%%%%%%%%%%%%%%%%%%%%%%
\section{Data and Sample Selection}\label{sec:data}
In this section, we describe the data set and sample construction used to investigate the Coma cluster by combining robust spectroscopic cluster membership with consistent galaxy physical, photometric, and structural measurements.

We adopt the Coma cluster properties from the $\mathtt{GalWCat19}$\footnote{\url{http://cdsarc.u-strasbg.fr/viz-bin/cat/J/ApJS/246/2}} galaxy cluster catalog \citep{GalWCat19Vizier}, constructed by applying the FoG-GalWeight toolkit to SDSS DR13 spectroscopic data. In this catalog, clusters are ranked by mass, and Coma corresponds to cluster number 3. The cluster center is located at $\alpha_c = 194.94^\circ$, $\delta_c = 27.91^\circ$, and redshift $z_c = 0.0234$, corresponding to a comoving distance of $D_c = 69.70~h^{-1}\,\mathrm{Mpc}$.

Cluster membership and projected phase-space boundaries are determined using the GalWeight technique \citep{Abdullah18}. Briefly, GalWeight assigns a weight to each galaxy according to its location in projected phase space, defined by its projected cluster-centric radius, $R_p$, and line-of-sight peculiar velocity, $V_{\mathrm{pec}}$. The total weight is defined as
\begin{equation}
\mathcal{W}_{\rm tot}(R_p,V_{\mathrm{pec}}) = \mathcal{W}_{\rm dy}(R_p,V_{\mathrm{pec}}) \mathcal{W}_{\rm ph}(R_p,V_{\mathrm{pec}}),
\end{equation}
where $\mathcal{W}_{\rm dy}$ and $\mathcal{W}_{\rm ph}$ are the dynamical and phase-space weights, respectively. The dynamical weight incorporates information from the projected radial and line-of-sight velocity distributions of galaxies, while the phase-space weight is derived from the galaxy density distribution in the $(R_p,V_{\mathrm{pec}})$ plane using a two-dimensional adaptive kernel estimator. The cluster membership boundary is then determined using the NDM (Number Density Method; \citealt{Abdullah13}), allowing probable cluster members to be separated from foreground and background interlopers. Further details and validation of the GalWeight technique are presented in \citet{Abdullah18}.

Global dynamical quantities are adopted directly from $\mathtt{GalWCat19}$, where they are derived using the virial mass estimator (e.g., \citealp{Binney87,Abdullah11}) and an NFW mass profile \citep{NFW96,NFW97}. The Coma cluster has a virial radius of $R_{200} = 1.54~h^{-1}\,\mathrm{Mpc}$, a velocity dispersion of $\sigma_{200} = 933^{+48}_{-55}~\mathrm{km~s^{-1}}$, and a virial mass of $M_{200} = (8.76 \pm 2.23) \times 10^{14}~h^{-1}\,M_\odot$. We also adopt a turnaround radius of $R_{\mathrm{ta}} = 6.77~h^{-1}\,\mathrm{Mpc}$, defined as the radius enclosing a mean density of $5.55\,\rho_c$ \citep{Nagamine03,Busha05,Dunner06}.

To characterize galaxy populations, we cross-match confirmed Coma members with two SDSS-based value-added catalogs. Stellar masses ($M_\ast$), star formation rates (SFRs), and specific star formation rates (sSFRs) are taken from the MPA/JHU catalog. The stellar masses were calculated following the methodologies described in \citet{Kauffmann03} and \citet{Salim07}, using the \citet{Bruzual03} stellar population synthesis models and assuming a \citet{Kroupa01} initial mass function. SFRs were calculated using the nebular emission lines within the 3 arcsec SDSS spectroscopic fiber following \citet{Brinchmann04}, while star formation outside the fiber aperture is estimated from galaxy photometry as described by \citet{Salim07}.  SFR estimates were primarily based on empirical $\mathrm{H}_{\alpha}$ calibrations \citep{Kennicutt98} and are corrected for dust extinction using the Balmer decrement, $\mathrm{H}_{\alpha}/\mathrm{H}_{\beta}$, following \citet{Charlot00}.
Structural and photometric parameters are obtained from \citet{Meert15}, which is based on refined two-dimensional surface brightness profile fitting and incorporates improved background subtraction techniques \citep{Vikram10,Bernardi13,Meert13,Bernardi14}. We apply quality cuts using the \texttt{finalflag} bitmask to exclude unreliable structural fits. Rest-frame $(g-r)$ colors are computed using extinction- and $K$-corrected photometry, and galaxy morphology is characterized using bulge-to-total ratios ($B/T$) derived from bulge--disk decompositions.

These catalogs are combined to construct a unified data set containing consistent dynamical, stellar population, photometric, and structural measurements for Coma galaxies. After cross-matching confirmed Coma members with the MPA/JHU catalog, the final sample contains 818 galaxies with stellar masses, SFRs, and sSFRs. Reliable structural measurements from \citet{Meert15}, including bulge-to-total ratios, are available for 738 of these galaxies after applying the adopted quality cuts. Therefore, we use the full 818-galaxy sample for analyses based on stellar mass, SFR, sSFR, and color, while analyses involving morphology or $B/T$ are restricted to the 738 galaxies with reliable structural measurements. 
The catalog of 818 galaxies is available as an electronic data product accompanying this article.

For operational classification, we divide galaxies according to star formation activity, color, and morphology using specific star formation rate, offset from the red sequence, and bulge-to-total ratio, respectively. We adopt fiducial thresholds of $\log(\mathrm{sSFR}/\mathrm{yr}^{-1}) = -10.5$, $\Delta(g-r)_{\rm RS} = -0.05$, and $B/T = 0.35$ to separate quiescent/star-forming, red/blue, and bulge/disk populations. These thresholds are used throughout this work as baseline definitions, while their empirical distributions and physical motivations are examined in Sections~\ref{sec:cmd}, \ref{sec:sfr} and Table~\ref{tab:pop_definitions}.
%%%%%%%%%%%%%%%%%%%%%%%%%%%%%%%%%%%%%%%%%%%%%%%%%%%%%%
\section{Results} \label{sec:results}

\subsection{Spatial Distribution and Projected Phase-Space Structure} \label{sec:dist}

We begin by examining the spatial and dynamical structure of galaxies in the Coma cluster region. Figure~\ref{fig:sky} (left panel) shows the sky distribution of galaxies centered on Coma. The galaxy distribution reveals a prominent overdensity associated with the cluster core, surrounded by a more diffuse population tracing the surrounding large-scale structure. The dashed circles indicate the characteristic radii of the system, $R_{200}$, $R_{\mathrm{sp}}$, and $R_{\mathrm{ta}}$, moving outward from the cluster center. These scales define the principal dynamical regimes of the cluster environment, corresponding to the virialized region, the splashback boundary, and the infall regime. The splashback radius, $R_{\mathrm{sp}}$, is measured independently from the surface density profile and is discussed in Section~\ref{sec:density}.

The middle panel presents the projected phase-space diagram, showing line-of-sight peculiar velocity ($V_{\mathrm{pec}}$) as a function of projected cluster-centric distance ($R_\mathrm{p}$). The characteristic trumpet-shaped structure is clearly visible \citep[e.g.,][]{Praton94, Abdullah13}, reflecting the gravitational potential of the cluster and the transition from the virialized core to the infall region. The solid curves trace the FoG-GalWeight boundaries used to assign cluster membership \citep{Abdullah18}, while the vertical dashed lines mark $R_{200}$, $R_{\mathrm{sp}}$, and $R_{\mathrm{ta}}$. Galaxies in the inner regions exhibit a broad velocity distribution, consistent with a virialized population, whereas at larger radii the velocity envelope progressively narrows, tracing recently accreted and infalling systems.

The right panel shows a slice of the Coma region in sky position and radial velocity, projected along the declination direction, providing a direct view of the cluster in redshift space. A prominent Finger-of-God (FoG) feature \citep{Jackson72,Kaiser87} dominates the central region, appearing as a vertically elongated structure produced by the large internal velocity dispersion of galaxies within the cluster core. This FoG structure is a signature of massive clusters in redshift space and directly reflects the deep gravitational potential of Coma. The virialized galaxy population is concentrated near the cluster center and exhibits a relatively symmetric velocity distribution about the cluster mean, consistent with a dynamically evolved core. In contrast, galaxies at larger projected radii show a more extended spatial distribution and narrower velocity structure, consistent with infalling populations that have not yet been fully virialized.

Overall, the spatial, projected phase-space, and redshift-space structure of Coma reveal a dynamically evolved core embedded within a continuously accreting outer environment. This clear separation between virialized and infall populations provides the physical framework for the remainder of this study. In particular, these dynamical boundaries establish the environmental architecture within which we interpret the transitions in galaxy star formation activity, stellar populations, and morphology from the cluster core to the splashback boundary and beyond.

%%%%%%%%%%%%%%%%%%%%%%%%%%%%%%%%%%%%%%%%%%%%%%%%%%%
%Figure 2_______________________________
\begin{figure*}\hspace{0cm}
\centering
\includegraphics[width=1\linewidth]{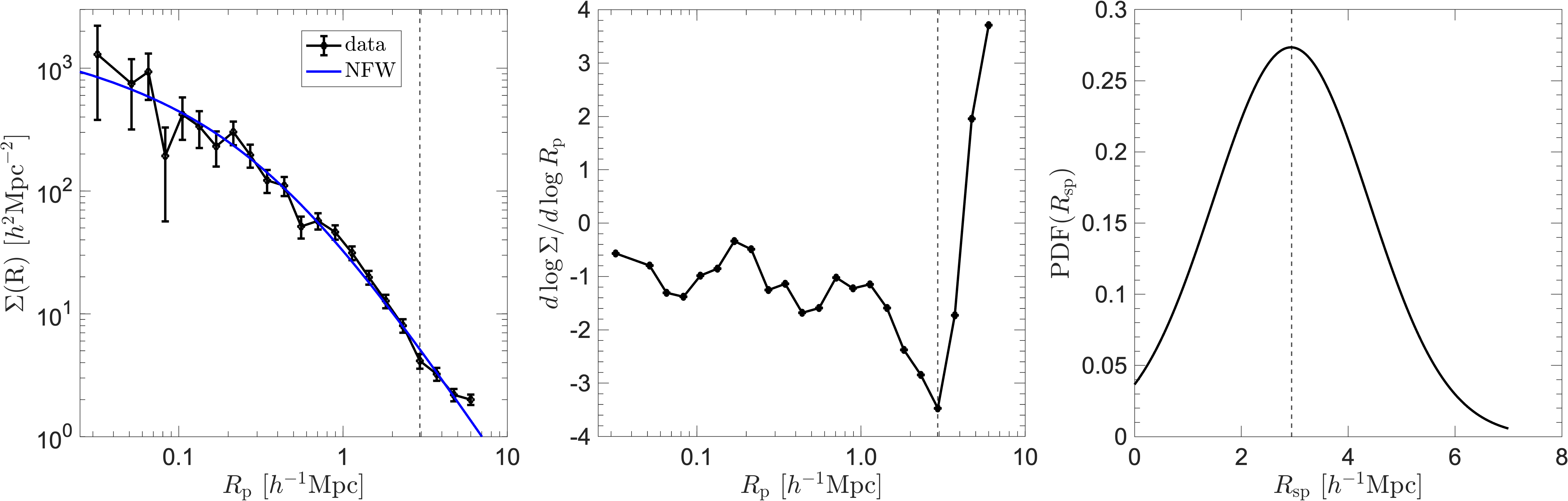} \vspace{-0.5cm}
    \caption{Surface number density profile and splashback radius of the Coma cluster. Left: projected surface number density profile $\Sigma(R_p)$ of cluster galaxies (points) along with the best-fit NFW model (blue line). Middle: logarithmic slope $d\ln\Sigma/d\ln R_{\rm p}$ as a function of projected radius. A clear minimum is observed at $R_{\mathrm{sp}}$, marking the splashback radius. Right: bootstrap distribution of $R_{\mathrm{sp}}$, showing a well-defined and unimodal peak, indicating a robust measurement. Vertical dashed lines indicate the inferred $R_{\mathrm{sp}}$ at 2.94~\h.}
    \label{fig:density}
\end{figure*}
%%%%%%%%%%%%%%%%%%%%%%%%%%%%%%%%%%%%%%%%%%%%%%%%%%%
\subsection{Surface Density Profile, Concentration, and Splashback Radius}
\label{sec:density}

In this section, we investigate the radial structure of the Coma cluster using the projected surface number density profile of member galaxies. By modeling this profile with an NFW model \citep{NFW96,NFW97}, we estimate the scale radius $r_s$ and concentration $c$, and then use the logarithmic slope of the profile to identify the splashback radius, $R_{\mathrm{sp}}$. This radius marks the transition between the virialized cluster interior and recently accreted material, providing a physically motivated structural boundary for the cluster.

The projected surface density profile, $\Sigma(R_p)$, is shown in Figure~\ref{fig:density} (left panel). The profile declines smoothly with radius over more than two orders of magnitude from the cluster core to the outskirts. We model this distribution using the Navarro--Frenk--White (NFW) density profile
\begin{equation}
\rho(r) = \frac{\rho_s}{x(1+x)^2},
\end{equation}
where $x=r/r_s$, $r_s$ is the scale radius, and $\rho_s=\delta_s\rho_c$ is the characteristic density within $r_s$. Here, $\delta_s = (\Delta_{200}/3) c^3 \left[\ln(1+c) - \frac{c}{1+c}\right]^{-1}$, and the concentration is defined as $c = R_{200}/r_s$ (e.g., \citealp{Mamon13}). The projected cumulative galaxy distribution within a cylinder of radius $R$ is given by \citep[e.g.,][]{Bartelmann96,Zenteno16}

\begin{equation}
\label{eq:NFW2}
N(<R) = \frac{N_s}{\ln(2) - \tfrac{1}{2}}
\begin{cases}
\ln\left(\frac{x}{2}\right) + \dfrac{\cosh^{-1}(1/x)}{\sqrt{1 - x^2}}
& \text{if } x < 1 \\[6pt]
1 - \ln(2)
& \text{if } x = 1 \\[6pt]
\ln\left(\frac{x}{2}\right) + \dfrac{\cos^{-1}(1/x)}{\sqrt{x^2 - 1}}
& \text{if } x > 1
\end{cases}.
\end{equation}
where $N_s$ is the normalization corresponding to the number of galaxies enclosed within $r_s$.
%%%%%%%%%%%%%%%%%%%%%%%%%%%%%%%%%%%%%%%%%%%%%%%%%%%%%%%%%%%%%
%Table 1_______________________________
\begin{table}
\centering
\caption{Derived structural and dynamical parameters of the Coma cluster}
\label{tab:coma_params}
\begin{tabular}{lcc}
\hline
\hline
Parameter & Value & Units \\
\hline
\multicolumn{3}{l}{\textit{Cluster Center}} \\
$\alpha_c$ & 194.935 & deg \\
$\delta_c$ & 27.912 & deg \\
$z_c$ & 0.0234 & -- \\
\hline
\multicolumn{3}{l}{\textit{Dynamical Properties}} \\
$R_{200}$ & 1.54 & $h^{-1}\,\mathrm{Mpc}$ \\
$\sigma_{200}$ & $933^{+48}_{-55}$ & $\mathrm{km\,s^{-1}}$ \\
$M_{200}$ & $(8.76 \pm 2.23)\times10^{14}$ & $h^{-1}\,M_\odot$ \\
$R_{\mathrm{ta}}$ & 6.77 & $h^{-1}\,\mathrm{Mpc}$ \\
\hline
\multicolumn{3}{l}{\textit{NFW Fit}} \\
$r_s$ & $0.42 \pm 0.28$ & $h^{-1}\,\mathrm{Mpc}$ \\
$c$ & 3.68 & -- \\
\hline
\multicolumn{3}{l}{\textit{Splashback}} \\
$R_{\mathrm{sp}}$ & $2.94 \pm 0.16$ & $h^{-1}\,\mathrm{Mpc}$ \\
$R_{\mathrm{sp}}/R_{200}$ & 1.91 & -- \\

\hline
\end{tabular}
\begin{tablenotes}
\item Dynamical parameters are taken from the FoG-GalWeight catalog \citep{Abdullah20a}. 
The uncertainty in $r_s$ is estimated from the $1\sigma$ likelihood interval, while the uncertainty in $R_{\rm sp}$ is estimated using bootstrap resampling.
\end{tablenotes}
\end{table}
%%%%%%%%%%%%%%%%%%%%%%%%%%%%%%%%%%%%%%%%%%%%%%%%%%%%%%%%%%%%%

%%%%%%%%%%%%%%%%%%%%%%%%%%%%%%%%%%%%%%%%%%%%%%%%%%%%%%%%%%%%%%%%%%%%%%%
%Figure 3_______________________________
\begin{figure*}\hspace{0cm}
\centering
\includegraphics[width=1\linewidth]{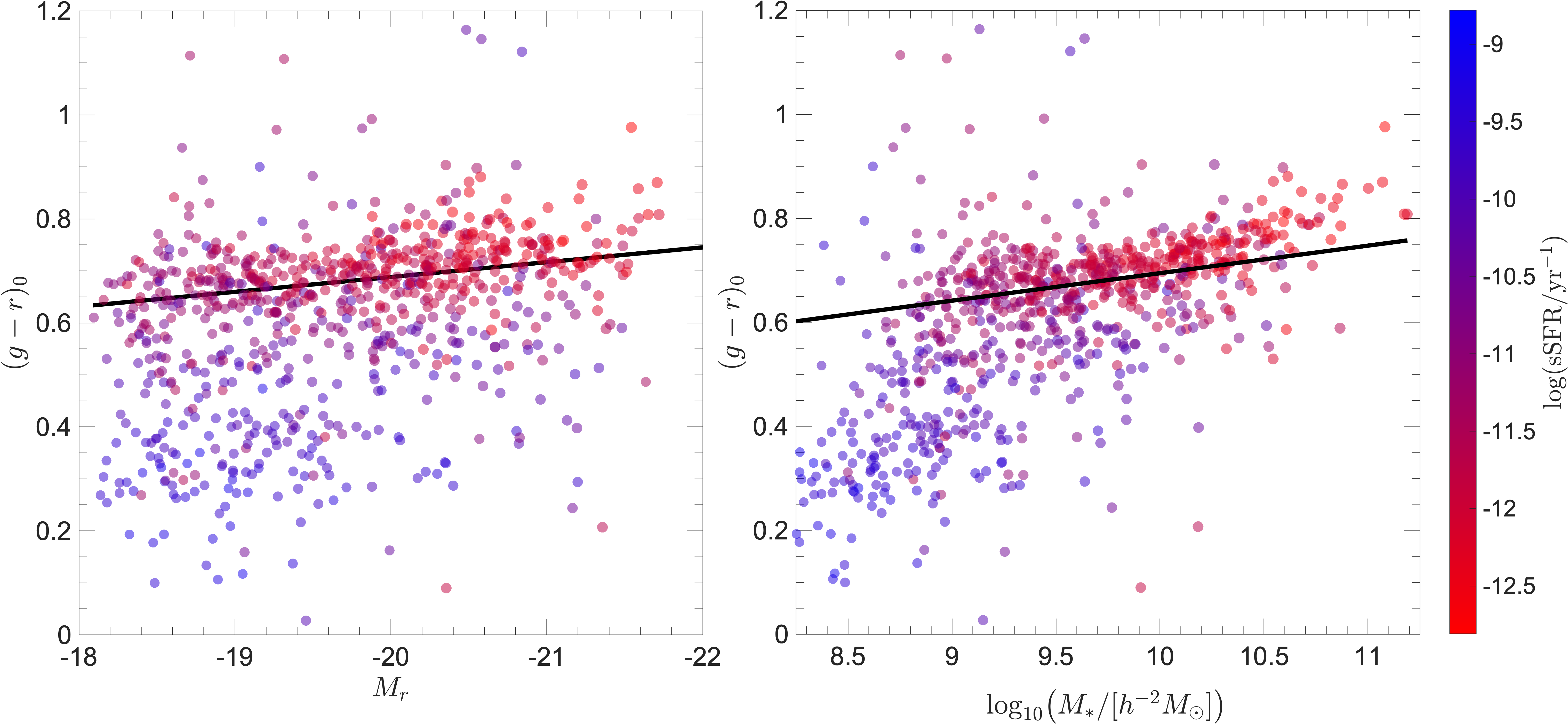} \vspace{-0.5cm}
\caption{
Color–magnitude (left) and color–stellar mass (right) diagrams for Coma cluster member galaxies. Points are color-coded by $\log(\mathrm{sSFR}/\mathrm{yr}^{-1})$, as indicated by the color bar. The black solid line represents the best-fit red sequence. The diagrams exhibit the characteristic bimodal distribution of galaxies into a tight red sequence and a more dispersed blue cloud, with an intermediate population occupying the transition region.
}
\label{fig:CMD}
\end{figure*}
%%%%%%%%%%%%%%%%%%%%%%%%%%%%%%%%%%%%%%%%%%%%%%%%%%%%%%%%%%%%%%%%%%%%%%%
Using the projected radii of Coma galaxies, we determine the NFW scale
radius through maximum-likelihood estimation by minimizing
\begin{equation}
-\ln\mathcal{L} = -\sum_i\ln\left[\frac{x_i\Sigma(x_i)} {\int_0^{x_{\max}}x\Sigma(x)\,dx}\right],
\end{equation}
where $x_i=R_{p,i}/r_s$, $x_{\max}=R_{\max}/r_s$, and $R_{\max}$ is the outer projected radius included in the fit.  Because the inferred NFW scale radius depends on the adopted outer fitting radius, we vary $R_{\max}$ from $0.5\,h^{-1}\,\mathrm{Mpc}$ to $3R_{200}$ in steps of $0.1\,h^{-1}\,\mathrm{Mpc}$. For each radial cut, we determine the best-fitting $r_s$ and then adopt the minimum value as the fiducial solution, following \citet{Abdullah20a}. We estimate the $1\sigma$ confidence interval using
$-\ln\mathcal{L}=-\ln\mathcal{L}_{\mathrm{ML}}+0.5$ \citep[e.g.,][]{Koranyi00,Mamon10,Mamon13}. From this analysis, we obtain an NFW scale radius of $r_s=0.42\pm0.28\,h^{-1}\,\mathrm{Mpc}$ and a corresponding
concentration parameter of $c=3.68$.

To identify the splashback radius, we smooth the surface number-density profile using a Savitzky--Golay filter. We then calculate its logarithmic slope, $d\ln\Sigma/d\ln R_{\rm p}$, shown in Figure~\ref{fig:density} (middle panel). The profile exhibits a clear minimum, corresponding to the steepest decline in galaxy density, which we identify as $R_{\rm sp}$. Its uncertainty is estimated using bootstrap resampling.
We assess the robustness of this measurement through bootstrap resampling of the galaxy catalog and repeated profile reconstruction. The resulting distribution of $R_{\rm sp}$ (Figure~\ref{fig:density}, right panel) shows a well-defined peak, indicating that the measurement is stable under bootstrap resampling of the galaxy catalog. From this procedure, we measure a splashback radius of $R_{\mathrm{sp}} = 2.94 \pm 0.16~h^{-1}\,\mathrm{Mpc}$, where the uncertainty is derived from the standard deviation of the bootstrap distribution.
This measurement establishes $R_{\mathrm{sp}}$ as a robust structural boundary in Coma and provides a critical physical scale for interpreting galaxy evolution throughout the remainder of this work. Our estimate is in good agreement with the recent independent measurement of $R_{\mathrm{sp}} = 2.67~h^{-1}\,\mathrm{Mpc}$ reported by \citet{Silva25}, providing additional validation of our analysis. The derived structural and dynamical properties of the Coma Cluster, including its center, global dynamical quantities, NFW parameters, and splashback radius, are summarized in Table~\ref{tab:coma_params}.
%%%%%%%%%%%%%%%%%%%%%%%%%%%%%%%%%%%%%%%%%%%%%%%%%%%%%%%%%%%%%%%%%%%%%%%
\subsection{Color--Magnitude and Color--Stellar Mass Diagrams}
\label{sec:cmd}

Figure~\ref{fig:CMD} shows the rest-frame $(g-r)$ color as a function of absolute magnitude (CMD; left panel) and stellar mass (CSMD; right panel) for Coma cluster member galaxies. Both diagrams exhibit the classical bimodal galaxy distribution, consisting of a tight red sequence, a broader blue cloud, and an intermediate transition population. This bimodality establishes the fundamental population structure of Coma and provides an important baseline for interpreting the more detailed environmental and multidimensional analyses presented in Sections~\ref{sec:sfr} and ~\ref{sec:joint}. The red sequence is well defined across a broad range of magnitudes and stellar masses, indicating a dominant population of quiescent galaxies with older stellar populations \citep[e.g.,][]{Terlevich01,Blanton09}. Its relatively small color scatter reflects a homogeneous population that has largely ceased significant star formation \citep[e.g.,][]{Bower92,Conroy14}. In contrast, the blue cloud consists of galaxies with ongoing or recent star formation and exhibits a broader color distribution \citep[e.g.,][]{Baldry04,Taylor15}. The broader spread of this population reflects greater diversity in stellar populations, gas content, and evolutionary state.

The CSMD further reveals a strong stellar mass dependence in galaxy populations. Red-sequence galaxies predominantly occupy the high-stellar-mass regime, while blue galaxies are more common at lower stellar masses. This mass-dependent color segregation reflects the close relationship between stellar mass growth, quenching, and galaxy evolution \citep[e.g.,][]{Kauffmann03,Peng10,Taylor15}. The color coding by sSFR further emphasizes the physical distinction between these populations, with red-sequence galaxies generally exhibiting systematically lower sSFRs than blue-cloud systems. Between the red sequence and blue cloud lies a lower-density intermediate bridge  in color space. This bridge corresponds to the region commonly referred to as the  green valley in color--magnitude and color--stellar mass diagrams, and is generally  interpreted as a transitional population of galaxies moving from active star formation toward quiescence \citep[e.g.,][]{Poggianti04}.

To quantify the red sequence, we define its ridge line in both the CMD and CSMD as the locus of maximum probability density in color at fixed magnitude or stellar mass using the two-dimensional adaptive kernel density estimate. These relations are modeled as
\begin{equation} \label{eq:color}
(g - r)_{\mathrm{RS}} = -0.029 \, M_r + 0.118,
\end{equation}
\begin{equation}
(g - r)_{\mathrm{RS}} = 0.053 \, \log(M_*) + 0.166.
\end{equation}

%%%%%%%%%%%%%%%%%%%%%%%%%%%%%%%%%%%%%%%%%%%%%%%%%%%%%%%%%%%%%%%%%%%%%%%
%Figure 4_______________________________
\begin{figure*}\hspace{0cm}
\centering
\includegraphics[width=0.9\linewidth]{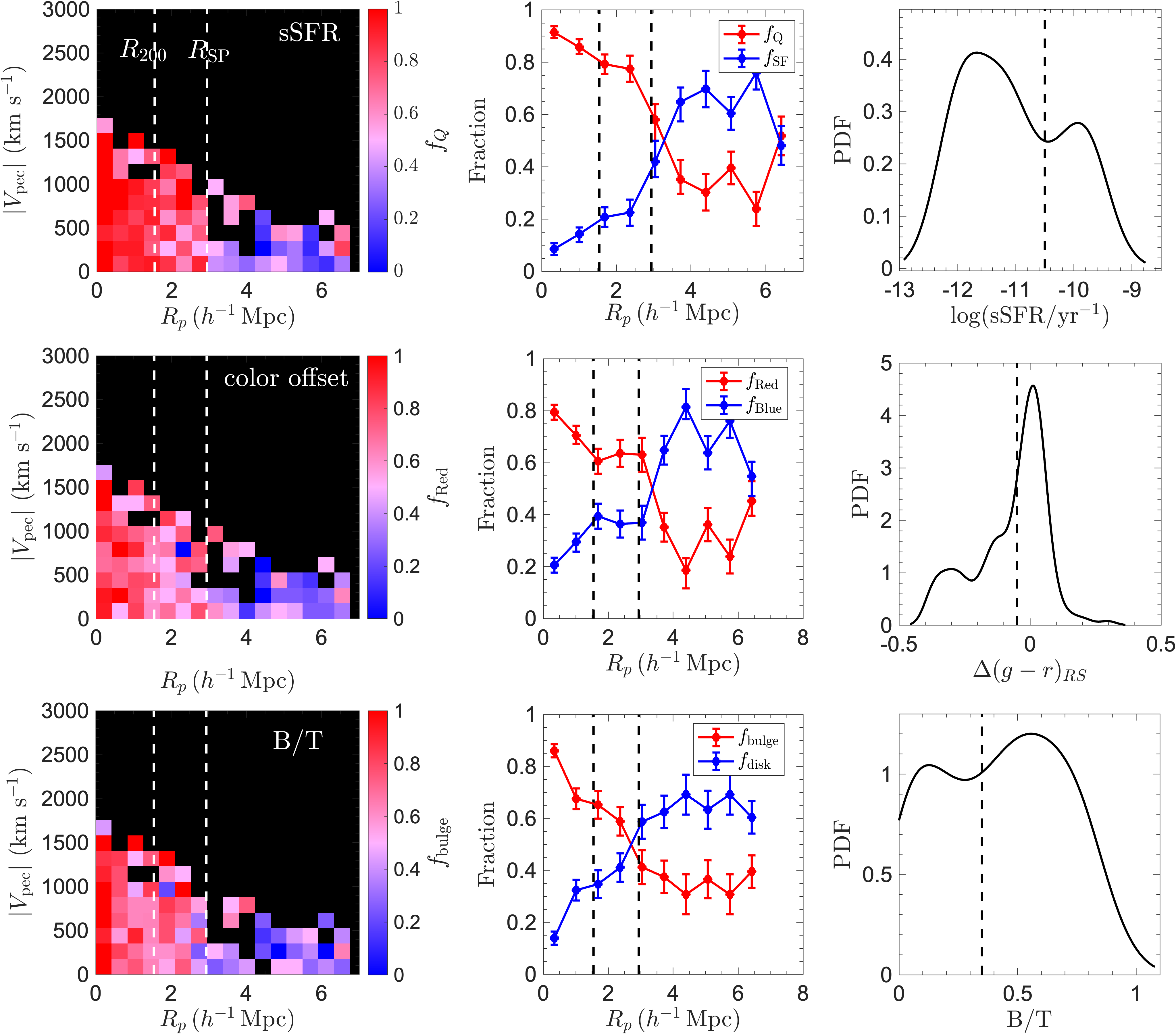} \vspace{-0.25cm}
\caption{
Projected phase-space distributions, radial population trends, and intrinsic property distributions of Coma cluster galaxies based on three independent galaxy population diagnostics: specific star formation rate (sSFR; top row), color offset from the red sequence $\Delta(g-r)_{\mathrm{RS}}$ (middle row), and bulge-to-total ratio ($B/T$; bottom row).
Left panels: two-dimensional maps in projected phase space showing the population fraction in each bin, with the color scale ranging from 0 to 1. The three rows show the fractions of quiescent galaxies ($f_{\rm Q}$), red-sequence galaxies ($f_{\rm Red}$), and bulge-dominated galaxies ($f_{\rm bulge}$), respectively, as functions of projected radius ($R_p$) and line-of-sight peculiar velocity ($|V_{\rm pec}|$).
Middle panels: radial population fractions measured in projected-radius bins, showing the corresponding complementary populations: star-forming, blue, and disk-dominated galaxies. Error bars are estimated via bootstrap resampling.
Right panels: one-dimensional probability density functions (PDFs) of sSFR, $\Delta(g-r)_{\mathrm{RS}}$, and $B/T$, illustrating the intrinsic population distributions used to define the adopted classification thresholds.
The vertical dashed lines in the left and middle panels mark the virial radius ($R_{200}$) and splashback radius ($R_{\mathrm{sp}}$), which are labeled directly on the figure.
The dashed lines in the right panels correspond to the fiducial divisions at $\log(\mathrm{sSFR}/\mathrm{yr}^{-1}) = -10.5$, $\Delta(g-r)_{\mathrm{RS}} = -0.05$, and $B/T = 0.35$, respectively.
}
\label{fig:fraction}
\end{figure*}
%%%%%%%%%%%%%%%%%%%%%%%%%%%%%%%%%%%%%%%%%%%%%%%%%%%%%%%%%%%%%%%%%%%%%%%\
Throughout this work, the color offset is calculated using the color--magnitude relation in Equation~\ref{eq:color}. For each galaxy, the offset is the difference between its observed color and the red-sequence color predicted at its absolute magnitude $M_r$. Positive offsets indicate galaxies redder than the red sequence, while negative offsets indicate bluer galaxies. The color--stellar-mass relation is shown for comparison and is not used to calculate the offset. This color--magnitude relation provides the baseline used to classify galaxy populations and connect traditional color bimodality to the broader environmental and three-dimensional evolutionary framework developed in Section~\ref{sec:joint}.
%%%%%%%%%%%%%%%%%%%%%%%%%%%%%%%%%%%%%%%%%%%%%%%%%%%%%%%%%%%%%%%%%%%%%%%
\subsection{Environmental Dependence of Star Formation, Color, and Morphology in Projected Phase Space}
\label{sec:sfr}

To investigate how galaxy evolution is linked to the dynamical structure of the Coma cluster, we examine three complementary tracers of galaxy evolutionary state: star formation activity, stellar population color, and morphology. Specifically, we use sSFR, color offset from the red sequence $\Delta(g-r)_{\mathrm{RS}}$, and bulge-to-total ratio $B/T$ to test whether galaxy transformation follows a coherent pattern across the cluster environment.

Figure~\ref{fig:fraction} presents the projected phase-space distribution of Coma galaxies as a function of $R_p$ and $|V_{\mathrm{pec}}|$ for each diagnostic independently. The left panels reveal a consistent environmental structure across all three properties. In the inner cluster region ($R_p \lesssim R_{200}$), galaxies are predominantly quiescent, red or close to the red sequence, and bulge-dominated. Toward larger radii, the fractions of star-forming, blue, and disk-dominated galaxies increase. This segregation indicates that the virialized cluster core is dominated by environmentally processed systems, while the outskirts and infall region contain a larger fraction of less evolved galaxies. These trends are not purely radial, but are also tied to the cluster dynamical structure, with the highest fractions of quiescent, red, and bulge-dominated galaxies concentrated in the dynamically hottest central regions.

The middle panels quantify these trends using radial population fractions measured in projected-radius bins. Across all three diagnostics, the quiescent, red-sequence, and bulge-dominated fractions decline systematically with increasing radius, while the star-forming, blue, and disk-dominated fractions rise. This behavior reveals a continuous environmental progression from a virialized, evolved core to a more mixed and actively accreting outer population. A major transition occurs near the splashback radius, $R_{\mathrm{sp}}$, where the dominant galaxy population changes substantially. The alignment of this transition in sSFR, color, and morphology indicates that $R_{\mathrm{sp}}$ is not only a structural or dynamical boundary, but also an evolutionary transition zone linked to the accretion history of cluster galaxies.

\begin{table*}
\centering
\caption{Operational galaxy population definitions and sample sizes used in Figure~\ref{fig:fraction}.}
\label{tab:pop_definitions}
\begin{tabular}{lcccc}
\hline
Diagnostic & Population 1 & $N_1$ & Population 2 & $N_2$ \\
\hline
Star formation activity 
& Quiescent, $\log(\mathrm{sSFR}/\mathrm{yr}^{-1}) \leq -10.5$ 
& 564 
& Star-forming, $\log(\mathrm{sSFR}/\mathrm{yr}^{-1}) > -10.5$ 
& 254 \\

Color offset 
& Red, $\Delta(g-r)_{\mathrm{RS}} \geq -0.05$ 
& 476 
& Blue, $\Delta(g-r)_{\mathrm{RS}} < -0.05$ 
& 342 \\

Morphology 
& Bulge-dominated, $B/T \geq 0.35$ 
& 438 
& Disk-dominated, $B/T < 0.35$ 
& 300 \\
\hline
\end{tabular}
\tablecomments{
The sSFR and color classifications are based on the full sample of 818 Coma member galaxies with MPA/JHU measurements. 
The morphology classification is based on the subset of 738 galaxies with reliable $B/T$ measurements from \citet{Meert15}. 
Here, $N_1$ and $N_2$ give the number of galaxies in each class for the corresponding diagnostic.
}
\end{table*}

The right panels show the intrinsic probability density functions (PDFs) of sSFR, $\Delta(g-r)_{\mathrm{RS}}$, and $B/T$, which motivate the fiducial classification thresholds used in this work. The dashed lines indicate the adopted divisions between quiescent and star-forming galaxies ($\log(\mathrm{sSFR}/\mathrm{yr}^{-1}) = -10.5$), red and blue populations ($\Delta(g-r)_{\mathrm{RS}} = -0.05$), and bulge- and disk-dominated systems ($B/T = 0.35$). These operational definitions, together with the number of galaxies in each class, are summarized in Table~\ref{tab:pop_definitions}. The sSFR threshold approximately traces the minimum between the quiescent and star-forming peaks, while the color threshold selects galaxies lying below the red-sequence ridge line and separates red-sequence galaxies from bluer systems. The morphological threshold approximately corresponds to the minimum between the disk-dominated and bulge-dominated peaks in the $B/T$ distribution. The sSFR distribution shows the clearest bimodality, while the color-offset distribution also separates the dominant populations. In contrast, the broader $B/T$ distribution suggests that morphological transformation may proceed more gradually, or across a wider range of structural states, than star formation quenching or color evolution. These thresholds are therefore used as operational divisions for comparing broad galaxy populations across the cluster environment.

Overall, these results demonstrate that galaxy populations in Coma show a coherent environmental dependence across multiple independent tracers. The consistent transitions in star formation activity, color, and morphology, especially near $R_{\mathrm{sp}}$, show that galaxy evolution is closely coupled to the cluster dynamical structure. This establishes the splashback radius as both a dynamical boundary and a physically meaningful evolutionary transition zone.
%%%%%%%%%%%%%%%%%%%%%%%%%%%%%%%%%%%%%%%%%%%%%%%%%%%%%%%%%%%%%%%%%%%%%%%%%%%%%
%Figure 5_______________________________
\begin{figure*}\hspace{0cm}
\centering
\includegraphics[width=1\linewidth]{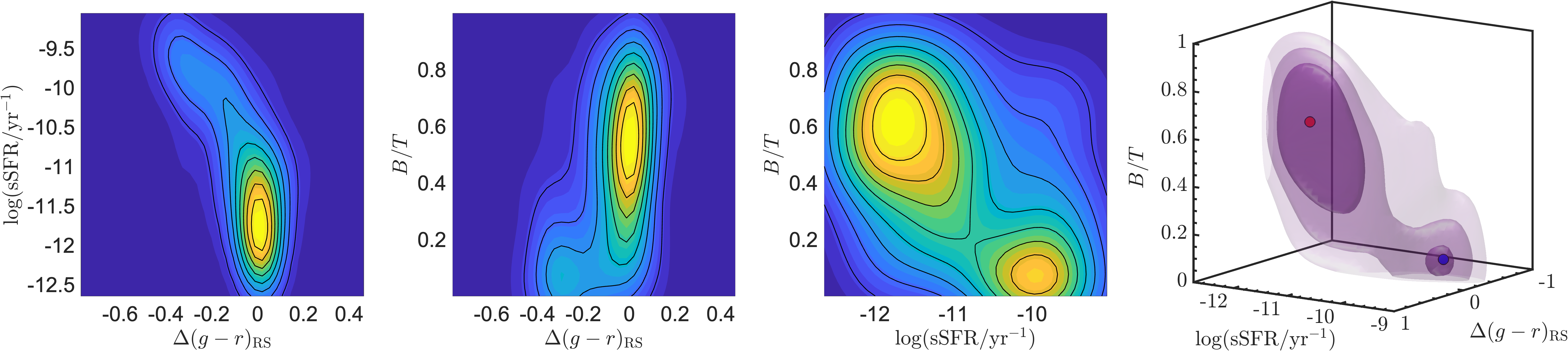} \vspace{-0.5cm}
\caption{
Joint distribution of galaxy properties in the Coma cluster based on a three-dimensional adaptive kernel map (3DAKM). The first three panels show the projected distributions onto the $(\log \mathrm{sSFR}, \Delta(g-r)_{\mathrm{RS}})$, $(B/T, \Delta(g-r)_{\mathrm{RS}})$, and $(B/T, \log \mathrm{sSFR})$ planes, respectively, obtained by marginalizing over the third parameter. 
The right panel presents the full three-dimensional distribution in the $(\log \mathrm{sSFR}, \Delta(g-r)_{\mathrm{RS}}, B/T)$ space. The red and blue points mark representative locations of quiescent and star-forming populations, respectively.
}
\label{fig:joint}
\end{figure*}
%%%%%%%%%%%%%%%%%%%%%%%%%%%%%%%%%%%%%%%%%%%%%%%%%%%%%%%%%%%%%%%%%%%%%%%%%%%%%
%Figure 6_______________________________
\begin{figure*}\hspace{0cm}
\centering
\includegraphics[width=1\linewidth]{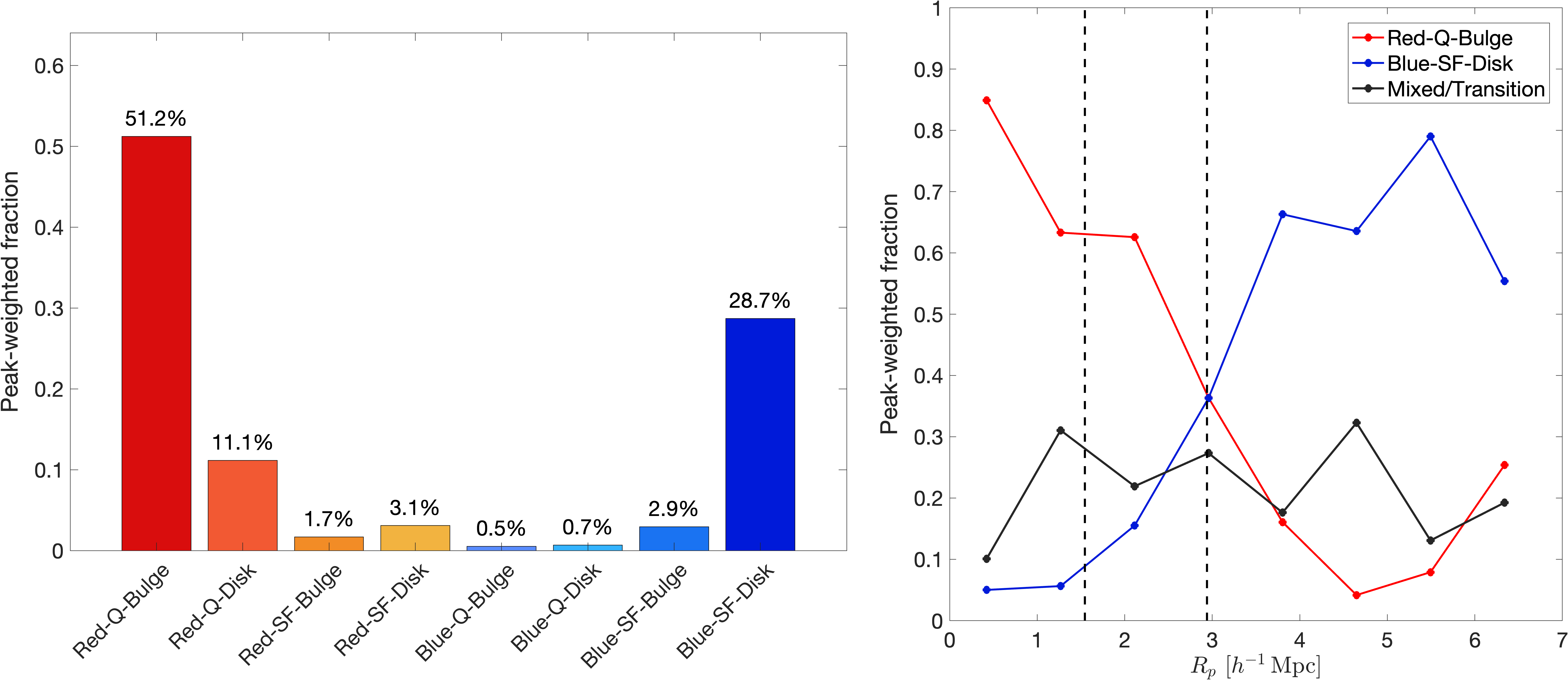} \vspace{-0.5cm}
\caption{
Peak-based probabilistic decomposition of the joint galaxy population in the Coma cluster using the 3DAKM in the $(\log \mathrm{sSFR}, \Delta(g-r)_{\mathrm{RS}}, B/T)$ space (Section~\ref{sec:joint} and Appendix~\ref{app:3dclass}). 
Left: global peak-weighted fractions of the eight possible joint galaxy classes, representing all combinations of red/blue, quiescent/star-forming, and bulge/disk properties. Right: radial variation of the three principal joint populations as a function of projected radius: the dominant red--quiescent--bulge population (red), the blue--star-forming--disk population (blue), and the combined mixed/transitional population (black), which includes all intermediate classes. The vertical dashed lines indicate the virial radius ($R_{200}$) and the splashback radius ($R_{\mathrm{sp}}$).
}
\label{fig:joint_frac}
\end{figure*}
%%%%%%%%%%%%%%%%%%%%%%%%%%%%%%%%%%%%%%%%%%%%%%%%%%%%%%%%%%%
\subsection{Joint Evolution of Star Formation, Color, and Morphology in the Coma Cluster}
\label{sec:joint}

Traditional galaxy classifications often divide systems into quiescent versus star-forming, red versus blue, or bulge- versus disk-dominated populations using sharp thresholds in individual properties such as $\log \mathrm{sSFR}$, $\Delta(g-r)_{\mathrm{RS}}$, or $B/T$. While effective for broad population studies, such one-dimensional divisions can artificially separate galaxies near classification boundaries and may obscure the intrinsically multidimensional nature of galaxy evolution, particularly in dense environments where star formation, stellar populations, and morphology are often shaped simultaneously but not necessarily transformed on identical timescales. To move beyond these limitations, we investigate the joint distribution of star formation activity, color, and morphology in order to determine whether Coma galaxies occupy distinct multidimensional evolutionary states and how these states connect to the cluster environment. {\bf{Because the joint analysis includes $B/T$, it is restricted to the 738 galaxies with reliable structural measurements.
}} 
Figure~\ref{fig:joint} presents the joint distribution of Coma galaxies in the three-dimensional $(\log \mathrm{sSFR}, \Delta(g-r)_{\mathrm{RS}}, B/T)$ parameter space using a three-dimensional adaptive kernel map (3DAKM). The projected distributions in each two-dimensional plane consistently reveal two dominant galaxy concentrations. Across all projections, the primary population corresponds to red, quiescent, bulge-dominated galaxies, while a secondary population corresponds to blue, star-forming, disk-dominated systems.

The full 3DAKM (Figure~\ref{fig:joint}, right panel) demonstrates that these are not merely pairwise correlations, but two physically distinct and coherent multidimensional evolutionary regions that dominate the Coma galaxy population. These two principal density maxima define the dominant evolutionary states occupied by galaxies in Coma: a red--quiescent--bulge-dominated state characterized by low sSFRs, positive $\Delta(g-r)_{\mathrm{RS}}$, and high $B/T$, and a blue--star-forming--disk-dominated state characterized by high sSFRs, negative $\Delta(g-r)_{\mathrm{RS}}$, and low $B/T$. This result demonstrates that star formation activity, color, and morphology are strongly coupled within the cluster environment, such that most galaxies occupy coherent multidimensional evolutionary basins rather than arbitrary combinations of galaxy properties.

Beyond these dominant endpoints, all projected planes and the full 3DAKM reveal a lower-density bridge connecting the two principal evolutionary regions. This bridge represents a substantial population of transitional or mixed systems, including quenched disks, red galaxies with residual star formation, and structurally evolving galaxies that do not fit neatly into classical categories. This intermediate manifold indicates that galaxy evolution in Coma proceeds through continuous multidimensional pathways, where star formation quenching, stellar population aging, and morphological transformation can occur on partially different timescales. For example, disk-dominated galaxies with low sSFR imply that star formation may be suppressed before full morphological transformation occurs, while red galaxies with relatively low $B/T$ indicate that stellar population aging or color evolution may precede or follow structural change. Thus, although galaxy properties evolve coherently within the cluster environment, they are shaped by multiple environmental processes operating at different rates.

To quantify these multidimensional evolutionary pathways, we next decompose the galaxy population into probabilistic joint classes based on the relative positions of galaxies to the two dominant 3DAKM peaks. Rather than applying hard thresholds in $\log \mathrm{sSFR}$, $\Delta(g-r)_{\mathrm{RS}}$, or $B/T$, we use the two principal multidimensional density maxima as evolutionary anchors. For each galaxy, peak-based probabilities are assigned independently along the star formation, color, and morphological axes according to its relative distance from the red--quiescent--bulge peak and the blue--star-forming--disk peak (see Appendix~\ref{app:3dclass}). This probabilistic framework preserves the continuous nature of galaxy evolution while enabling the galaxy population to be decomposed into physically meaningful multidimensional classes.

Using this framework, galaxies are decomposed into eight possible joint evolutionary classes representing all combinations of red/blue, quiescent/star-forming, and bulge/disk properties. Figure~\ref{fig:joint_frac} shows both the global population fractions (left panel) and their radial dependence (right panel). We find that Coma is dominated by two principal populations: red--quiescent--bulge-dominated galaxies, which constitute $51.2\%$ of the cluster population, and blue--star-forming--disk-dominated galaxies, which account for $\sim29\%$. The remaining $\sim20\%$ of galaxies occupy mixed or transitional evolutionary states, collectively tracing the multidimensional bridge between the two dominant populations. These mixed classes quantify the substantial intermediate manifold identified in the 3DAKM and indicate that galaxy transformation can proceed through physically meaningful transitional states such as quenched disks or morphologically transformed yet still star-forming systems.

The radial trends shown in the right panel of Figure~\ref{fig:joint_frac} demonstrate that these multidimensional classes are strongly linked to the dynamical structure of the cluster. The red--quiescent--bulge-dominated population dominates the inner cluster core, where it constitutes more than $80\%$ of galaxies at small projected radii. Its fraction declines steadily with radius, dropping sharply near $R_{200}$ and approaching parity with the blue--star-forming--disk population near the splashback radius $R_{\mathrm{sp}}$. Beyond this boundary, the blue--star-forming--disk class becomes dominant, consistent with a population of recently accreted galaxies that have not yet fully experienced cluster-driven transformation.

Most notably, the strongest shift in multidimensional population balance occurs near $R_{\mathrm{sp}}$, suggesting that this characteristic radius is not merely a dynamical boundary, but a major evolutionary transition zone where environmental processes such as ram-pressure stripping, starvation, harassment, and morphological restructuring are actively reshaping galaxies. In this picture, the splashback region marks the location where galaxies are most likely to move through intermediate evolutionary pathways before joining the dominant red cluster population.

Our results are consistent with previous studies of galaxy populations in Coma and nearby clusters. The increase of quiescent, red, and bulge-dominated galaxies toward the cluster center agrees with the classical morphology--density relation, in which the fractions of elliptical and S0 galaxies increase with local density, while the fraction of spiral galaxies decreases \citep{Dressler80}. This trend is also consistent with the morphology--radius relation, where early-type galaxies become increasingly dominant toward smaller cluster-centric radii \citep{Whitmore93}. For Coma specifically, \citet{Price11} found that most bright cluster members are passive or quiescent, with only a smaller fraction showing significant emission-line activity. UV and infrared studies of Coma further show that star-forming galaxies are preferentially associated with the outskirts, infall regions, or less dynamically evolved environments \citep{Cortese08,Mahajan10}. More recently, \citet{Upadhyay21} connected the star-formation histories of Coma galaxies to their orbital histories and suggested that quenching is closely linked to cluster infall and pericentric passage. In addition, the dominance of red and bulge-dominated systems agrees with structural studies of the Coma red sequence, where many galaxies are well described by bulge-plus-disc structures whose properties vary with cluster-centric radius \citep{Head14}. Therefore, our finding that the inner Coma region is dominated by quiescent, red, and bulge-dominated galaxies, while star-forming, blue, and disk-dominated systems become more common at larger projected radii, is consistent with previous observational studies. The main contribution of this work is that we connect these established trends using three independent diagnostics, sSFR, color offset from the red sequence, and bulge-to-total ratio, within a unified projected phase-space and radial framework.

%%%%%%%%%%%%%%%%%%%%%%%%%%%%%%%
\section{Conclusion} \label{sec:conc}

In this work, we present a comprehensive investigation of galaxy evolution across the full dynamical structure of the Coma cluster using the FoG-GalWeight spectroscopic cluster framework combined with SDSS-based value-added galaxy properties. By unifying robust cluster membership, projected phase-space structure, star formation activity, stellar populations, and morphology within a consistent framework, we examine how galaxy transformation proceeds from the virialized cluster core to the splashback boundary and infall region.

Our main results are summarized as follows:

\begin{enumerate}
\item We characterize the projected spatial and dynamical structure of Coma and measure a splashback radius of $R_{\mathrm{sp}} = 2.94 \pm 0.16~h^{-1}\,\mathrm{Mpc}$, corresponding to $R_{\mathrm{sp}}/R_{200} \approx 1.91$. This measurement is consistent with recent independent studies and identifies a physically meaningful dynamical boundary separating the virialized cluster interior from the infall region.

\item Using specific star formation rate, color offset from the red sequence, and bulge-to-total ratio independently, we find consistent environmental transitions from quiescent, red, bulge-dominated galaxies in the cluster core to increasingly star-forming, blue, disk-dominated populations toward larger cluster-centric radii. These transitions occur continuously across the cluster environment, with major population shifts occurring near the splashback radius.

\item We introduce a new three-dimensional framework in the joint $(\log \mathrm{sSFR}, \Delta(g-r)_{\mathrm{RS}}, B/T)$ space and develop a novel peak-based classification scheme that extends beyond traditional one-dimensional galaxy classifications. This framework reveals two dominant evolutionary populations: red, quiescent, bulge-dominated galaxies (51\%) and blue, star-forming, disk-dominated galaxies (29\%), together with a remaining $\sim$20\% population of transitional or mixed systems connecting these two principal states.

\item The relative fractions of these multidimensional galaxy populations vary strongly with environment. The red, quiescent, bulge-dominated population dominates the virialized cluster core, while the blue, star-forming, disk-dominated population becomes increasingly important toward larger radii. The strongest transition occurs near the splashback radius, demonstrating that $R_{\mathrm{sp}}$ is not only a structural or dynamical boundary, but also a critical evolutionary transition zone.

\item Our results show that galaxy evolution in Coma is not purely binary. Instead, galaxies occupy continuous multidimensional evolutionary pathways in which star formation quenching, stellar population aging, and morphological transformation are tightly coupled but can proceed on partially different timescales. The substantial transitional population identified in our analysis suggests that environmental transformation is progressive rather than instantaneous.
\end{enumerate}

Overall, this work establishes the Coma cluster as a benchmark for linking cluster dynamical structure to galaxy evolution. We show that its galaxy population is organized around two dominant evolutionary basins, broadly corresponding to star-forming disk-dominated and quiescent bulge-dominated systems, connected by a transitional population. By combining a peak-based multidimensional classification scheme with the cluster dynamical structure, particularly the splashback boundary, we show that galaxy transformation in dense environments is a continuous process regulated by environment, accretion history, and dynamical structure. This framework provides a physically motivated approach for studying galaxy evolution in other clusters and large spectroscopic surveys.

\section*{Acknowledgements}
GW gratefully acknowledges support from the National Science Foundation through grant AST-2347348.

\appendix
\section{Peak-Based Three-Dimensional Classification Framework}
\label{app:3dclass}

To quantify the multidimensional galaxy populations identified in Section~\ref{sec:joint}, we construct a peak-based probabilistic classification framework using the 3DAKM map in the joint $(\Delta(g-r)_{\mathrm{RS}}, \log \mathrm{sSFR}, B/T)$ parameter space. The 3DAKM shown in Figure~\ref{fig:joint} reveals two dominant multidimensional density maxima corresponding to the principal multidimensional evolutionary populations in Coma. We define these as the early ($E$) and late ($L$)
evolutionary states. The early peak corresponds to the
red, quiescent, bulge-dominated population, while the
late peak corresponds to the blue, star-forming, disk-dominated population. Along the color axis, the early and late anchors correspond to red and blue states, respectively; along the star-formation axis, they correspond to quiescent and star-forming states; and along the morphological axis, they correspond to bulge- and disk-dominated states.

These two peaks are defined by the primary maxima
of the smoothed 3DAKM density field:
\begin{equation}
\mathbf{X}_{\mathrm{E}} = (\Delta_{\mathrm{E}}, S_{\mathrm{E}}, BT_{\mathrm{E}})
\end{equation}
\begin{equation}
\mathbf{X}_{\mathrm{L}} = (\Delta_{\mathrm{L}}, S_{\mathrm{L}}, BT_{\mathrm{L}}),
\end{equation}
where $\Delta \equiv \Delta(g-r)_{\mathrm{RS}}$, $S \equiv \log \mathrm{sSFR}$, and $BT \equiv B/T$. These peak coordinates define the fiducial multidimensional evolutionary endpoints used as reference anchors throughout the classification framework.

Using these two evolutionary anchors, directional weights are constructed independently along the color, star formation, and morphological dimensions. For a given parameter $X \in \{\Delta, S, BT\}$ and evolutionary state $C \in \{E,L\}$, the directional weight for galaxy $i$ is defined as
\begin{equation}
W_{X,C}(i) =
\exp\left[
-\frac{1}{2}
\left(
\frac{X_i - X_C}{\sigma_X}
\right)^2
\right],
\end{equation}
where $X_i$ is the galaxy property value, $X_C$ is the corresponding peak coordinate, and $\sigma_X$ is the global standard deviation of that parameter measured from the full Coma galaxy population. We adopt global standard deviations to provide a uniform metric across the full galaxy population while preserving the dominant multidimensional topology without overfitting to local peak structure.

This yields six directional weights for each galaxy:
$(W_{\Delta,E}, W_{\Delta,L})$ for color,
$(W_{S,E}, W_{S,L})$ for star formation activity, and
$(W_{BT,E}, W_{BT,L})$ for morphology,
which quantify the galaxy’s relative proximity to the early and late evolutionary peaks along each dimension independently.

The joint weight for each multidimensional evolutionary class is then obtained by combining these directional weights across all three dimensions:
\begin{equation}
P_{C_{\Delta},C_{S},C_{BT}}(i)
=
W_{\Delta,C_{\Delta}}(i)\,
W_{S,C_{S}}(i)\,
W_{BT,C_{BT}}(i),
\end{equation}
where each of $C_{\Delta}$, $C_{S}$, and $C_{BT}$ can independently take either the early ($E$) or late ($L$) state. The eight possible combinations of $(C_{\Delta},C_{S},C_{BT})$ correspond directly to the eight joint evolutionary classes RQB, RQD, RSB, RSD, BQB, BQD, BSB, and BSD, where R/B denote red/blue color states, Q/S denote quiescent/star-forming states, and B/D denote bulge-/disk-dominated morphologies.

To convert the joint weights into normalized class probabilities for each galaxy, we define
\begin{equation}
\tilde{P}_k(i)
=
\frac{P_k(i)}
{\sum_{j=1}^{8} P_j(i)},
\end{equation}
such that
\begin{equation}
\sum_{k=1}^{8}\tilde{P}_k(i)=1.
\end{equation}

Rather than assigning galaxies exclusively to a single discrete class, this framework allows each galaxy to contribute probabilistically to all eight classes according to its normalized multidimensional proximity to the two evolutionary anchors. The global fraction of each class is therefore computed as
\begin{equation}
f_k = \frac{1}{N} \sum_{i=1}^{N}\tilde{P}_k(i),
\end{equation}
where $N$ is the total number of galaxies.

This framework avoids sharp one-dimensional boundaries and preserves the continuous nature of galaxy evolution by allowing galaxies near the bridge between the two dominant density peaks to contribute naturally to intermediate multidimensional classes. It therefore provides a physically motivated description of galaxy evolution in Coma as continuous occupation of multidimensional evolutionary pathways rather than a purely binary transformation. 

\bibliography{ref}{}
\bibliographystyle{aasjournal}
%%%%%%%%%%%%%%%%% APPENDICES %%%%%%%%%%%%%%%%%%%%%
\end{document}